\documentclass[showpacs,preprintnumbers,amsmath,amssymb]{revtex4}

\usepackage{graphicx}
\usepackage{subfigure}
\usepackage{dcolumn}
\usepackage{bm}

\begin{document}

\preprint{APS/123-QED}

\title{Diffusion in the Markovian limit of the spatio-temporal colored noise}

\author{Takaaki Monnai}%
\email{monnai@suou.waseda.jp}%
\author{Ayumu Sugita}
 \email{sugita@a-phys.eng.osaka-cu.ac.jp}
\author{Katsuhiro Nakamura}
\email{nakamura@a-phy.eng.osaka-cu.ac.jp}
\affiliation{$*$Department of Applied Physics ,Waseda University, 3-4-1 Okubo,
Shinjuku-ku, Tokyo 169-8555, Japan\\
$\dagger $$\ddagger $Department of Applied Physics, Osaka City University,
3-3-138 Sugimoto, Sumiyoshi-ku, Osaka 558-8585, Japan}

\date{\today}

\begin{abstract}
We explore the diffusion process in the non-Markovian spatio-temporal noise.
There is a non-trivial short memory regime, i.e., the Markovian limit characterized by a scaling relation between the spatial and temporal correlation lengths.
In this regime, a Fokker-Planck equation is derived by expanding the trajectory around the
 systematic motion and the non-Markovian nature amounts to the systematic reduction of the
potential. For a system with the potential barrier, this fact leads to the  renormalization of  both the barrier height and collisional prefactor in the Kramers escape rate, with the resultant rate showing a maximum at some scaling limit.
\end{abstract}

\pacs{05.40.-a, 05.60.Cd, 82.20.Db}
\maketitle
It has been a long-standing problem to systematically treat the spatio-temporal colored noise in the thermal diffusion\cite{Golbovech,Rosenbluth,Deutsch}.
This is in contrast to the successful development  of the noise-induced diffusion processes in the cases of  additive noise\cite{Risken}, multiplicative noise\cite{Sancho}, and temporally colored noise
\cite{Doering}.
In the context of path coalescence, Deutsch\cite{Deutsch} first examined the motion of a damped particle subjected to a force fluctuating in both space and time.  Wilkinson, Mehlig and coworkers
\cite{Arvedson,Wilkinson,Mehlig}  pursued a similar subject, intensively explored the generalized Ornstein-Uhlenbeck process in momentum space, and showed  the anomalous diffusion where
the  spatial memory led to the  staggered ladder spectra\cite{Arvedson}.
A typical example of the spatio-temporal noise is given by the turbulent flow or a randomly moving gas which suspends small particles,
and space coordinates often stand for collective or reaction coordinates and order parameters in general.
Then small particles or molecules suspended by the turbulent flow can experience systematic forces.
The above works\cite{Deutsch, Arvedson,Wilkinson,Mehlig}, however, are limited to the diffusion with no systematic force.
It is thus a natural challenge to consider the diffusion with a spatio-temporal correlated noise in the presence of the systematic force induced by a potential.
The motivation of the present paper is to elucidate a scenario how the spatio-temporal correlated noise may radically affect the potential-induced motion with the purely-temporal noise.
In this case, we shall see that there is a non-trivial Markovian limit characterized by the spatial and temporal correlation lengths,  and  the Fokker-Planck equation for the distribution of  position, collective coordinates or the order parameter in general, is derived from the conservation of population.
Then we shall investigate the Kramers escape rate problem where the potential renormalization plays a crucial role.
The new escape rate matches up to renewed attentions paid for the Kramers escape rate theory\cite{Hanngi,Monnai,Monnai2}.

This Letter is organized as follows:
After a sketch of our model, the Markovian limit is discussed and then the Fokker-Planck equation is derived from the conservation of the population.
The escape rate is derived as an inverse of the mean first passage time.

Our model is described by the overdamped Langevin equation with a  colored noise
\begin{eqnarray}
&&\eta\dot{x}(t)=-U'(x(t))+f(x(t),t)\nonumber \\
&&\langle f(x,t)\rangle =0, \langle f(x,t)f(x',t')\rangle =C(x-x',t-t'),
\label{basiceq}
\end{eqnarray}
where $\eta$ stands for the friction coefficient and the potential $U(x)$ consists of
a metastable well and a potential barrier (see Fig.1).
\begin{figure}
\center{
\includegraphics[scale=0.8]{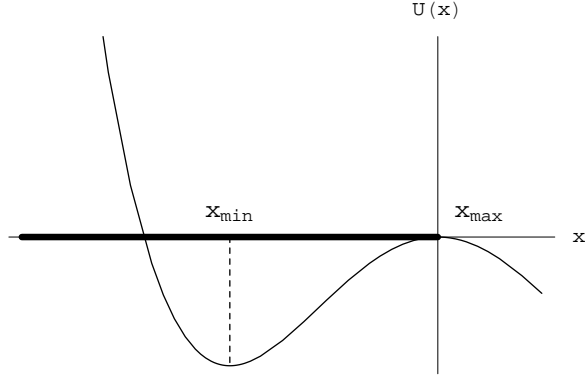}
}
\caption{The schematic illustration of the potential $U(x)$. The thick line indicates the domain of the metastable state.}
\label{metastable}
\end{figure}
The bracket $\langle\rangle$ stands for the average over the noise $f(x,t)$ and the initial distribution.
The noise $f(x,t)$ is assumed as a Gaussian process in accordance with the central limit theorem
both in the spatial and temporal variables.
An explicit expression of the noise $f(x,t)$ is given in the later paragraph concerning our stochastic simulation.
As a typical form of the autocorrelation function, we consider the Gaussian memory
\begin{equation}
C(x,t)=C_0 e^{-\frac{x^2}{\xi^2}-\frac{t^2}{\tau^2}},
\end{equation}
where $\xi$ is the spatial correlation length and $\tau$ is the correlation time, respectively.
The constant $C_0$ contains $\tau$ and $\xi$ in general.
Its asymptotic form is fully determined in a short correlation limit which we shall discuss below.
We note that essentially the same result is obtained for the other form of the memory, $C(x,t)=C_0 \frac{\sin(x/\xi)}{x}e^{-t^2/\tau^2}$, which 
behaves as Gaussian at the origin, and converges to the Dirac-delta. 
We assume that the usual overdamped Langevin equation is recovered as the position-independent noise case
by taking the limit of $\xi\rightarrow \infty$ and the temporal Markovian limit $\tau\rightarrow 0$ with $C_0=2\eta k_B T/\tau\sqrt{\pi}$.
There is still a possibility of $\xi$ dependence as a Laurent series
\begin{equation}
C_0\approx\frac{2\eta k_B T}{\sqrt{\pi}\tau}(\frac{C^{(0)}}{\xi^0}+\frac{C^{(1)}}{\xi}+\frac{C^{(2)}}{\xi^2}+..+\frac{C^{(n)}}{\xi^n}),
\label{correlation}
\end{equation}
with the strongest divergence $C^{(n)}/\xi^n, (n\geq 0)$ where
$C^{(n)}/\xi^n$ is dimensionless and $C^{(0)}= 1$.
It will become clear that all the terms should vanish $C^{(n)}=0$ for $n>0$, by requiring that  the diffusion coefficient is finite in the short correlation limit $\xi\rightarrow 0$.
At the moment, however, we employ the general expression (\ref{correlation}).
 Note that Sinai's random walk in random environment leads to the anomalously slow diffusion\cite{Sinai}, while
in our case symmetric nature of random force may enhance the diffusion over the barrier.

The Fokker-Planck equation at a given instant is obtained from the conservation of probability\cite{Zwanzig}.
The stochastic Liouville equation for the probability distribution $p(x,t)$ of each realization of noise is given as
\begin{equation}
\frac{\partial}{\partial t}p(x,t)=-L p(x,t)-\frac{1}{\eta}\frac{\partial}{\partial x}f(x,t)p(x,t),\label{stochastic}
\end{equation}
where $L p(x,t)=-\frac{\partial}{\partial x}\frac{1}{\eta}U'(x)p(x,t)$ is the deterministic Liouville operator.
With use of the identity
\begin{equation}
p(x,t)=e^{-tL}p(x,0)-\frac{1}{\eta}\int_0^t ds e^{(s-t)L}\frac{\partial}{\partial x}f(x,s)p(x,s),\label{probdist}
\end{equation}
the equation (\ref{stochastic}) is rewritten in the form of diffusion equation as
\begin{eqnarray}
&&\frac{\partial}{\partial t}p(x,t)=-L p(x,t)-\frac{\partial}{\partial x}\frac{1}{\eta}f(x,t)e^{-t L}p(x,0)\nonumber \\
&&+\frac{1}{\eta^2}\frac{\partial}{\partial x}f(x,t)\int_0^t ds e^{-(t-s)L}\frac{\partial}{\partial x}f(x,s)p(x,s).
\end{eqnarray}
By averaging over the noise, the population distribution $P(x,t)\equiv\langle p(x,t)\rangle$ obeys the diffusion equation,
\begin{equation}
\frac{\partial}{\partial t}P(x,t)
=\frac{1}{\eta}\frac{\partial}{\partial x}U'(x)P(x,t)+\frac{1}{\eta^2}\frac{\partial}{\partial x}
\int_0^t ds \langle f(x,t)\frac{\partial}{\partial x}f(y(x),s)p(y(x),s)\rangle.
\label{popudistr}
\end{equation}
Here, $y(x)\equiv e^{(t-s)L}x=x+\int_0^{t-s}ds' U'(x(s'))/\eta$ is the time-reversed deterministic trajectory which starts at $x$.
The last term in Eq. (\ref{popudistr}) is evaluated as follows.
After the averaging, we shall take the Markovian limit $\tau\rightarrow 0$, and thus the correlation between
 explicit noise terms and those at earlier times implicitly contained in the distribution $p(y(x),s)$ is neglected.
More precisely, the correlation between $f$ and $p$ is calculated by substituting the identity (\ref{probdist}) into the stochastic Liouville equation twice, and applying the pair-wise factorization for Gaussian variables:
\begin{eqnarray}
&&\int_0^t ds\int_0^s ds'\int_0^{s'}ds''1/\tau^2 e^{-(t-s')^2/\tau^2}e^{-(s-s'')^2/\tau^2}\nonumber \\
&=&O(\tau)  (\tau\rightarrow 0),
\end{eqnarray}
which turns out to be negligible in the Markovian limit.
On the other hand, the correlation between noise terms is
\begin{eqnarray}
&&\int_0^t ds\langle f(x,t)(\frac{\partial}{\partial x}f(y(x),s))\rangle\nonumber \\
&=&\int_0^t ds C_0\frac{dy}{dx}\frac{\partial}{\partial y(x)}e^{-\frac{(x-y(x))^2}{\xi^2}-\frac{(t-s)^2}{\tau^2}}|_{y=y(x)}\nonumber \\
&\approx&-\int_0^t ds C_0 (1+O(\tau))\frac{2 U'(x)}{\eta\xi^2}(t-s)e^{-(\frac{U'(x)^2}{\eta^2\xi^2}+\frac{1}{\tau^2})(t-s)^2}\nonumber \\
&\approx&- C_0\frac{U'(x)}{\eta\xi^2}\frac{\xi^2\eta^2}{U'(x)^2+\frac{\xi^2\eta^2}{\tau^2}}.
\label{renorm}
\end{eqnarray}
This means  a systematic renormalization of the potential.
We approximated the trajectory as $y(x)=x+\int_0^{t-s} ds'U'(x(s'))/\eta\approx x+U'(x)(t-s)/\eta$, because it appears
in the exponentially decaying factor as $t-s$ gets larger and, in the Markovian limit, only the short time integral does contribute to the correlation function.
Choosing the constant $C_0$ as Eq.(\ref{correlation}) so that the usual Langevin equation is recovered as a special case,
the dominant contribution of Eq. (\ref{renorm}) to the Fokker-Planck equation is
\begin{equation}
-U'(x)\frac{2C^{(n)}\eta k_B T}{\sqrt{\pi}(\tau\xi^n U'(x)^2+\frac{\xi^{n+2}\eta^2}{\tau})} \quad,
\end{equation}
where a Markovian limit is well-defined: Keeping the dimensionless parameter
\begin{equation}
\kappa_n=2C^{(n)}\tau k_B T/\sqrt{\pi}\xi^{n+2}\eta
\end{equation}
 as a constant,
both the temporal and spatial correlations go to zero,
$\tau, \xi\rightarrow 0$. Similarly the diffusion coefficient is given as
\begin{equation}
\frac{k_B T}{\eta}\frac{C^{(n)}}{\xi^n}\frac{\xi\eta}{\sqrt{(\tau U'(x))^2+(\xi\eta)^2}} \quad,
\end{equation}
which diverges in the limit  $\tau, \xi\rightarrow 0$ for $n>0$.
Thus hereafter we consider the case of $n=0$, where the diffusion coefficient approaches $\frac{k_B T}{\eta}$ in the Markovian limit specified by $\kappa\equiv\kappa_0$.
It is instructive to derive the diffusion constant in a more general way.
We assume that the kernel $C(x,t)$ rapidly vanishes as $t/\tau\rightarrow \infty$.
Then the time-integral of noise correlation function is calculated as
\begin{eqnarray}
&&\frac{1}{\eta^2}\int_0^t ds C(x-y(x),t-s)\nonumber \\
&=&\frac{1}{\eta^2}\int_0^t ds C(\xi\frac{x-y(x)}{\xi},\tau\frac{t-s}{\tau})\nonumber \\
&\approx&\frac{1}{\eta^2}\int_0^t ds C(\xi\frac{U'(x)\tau}{\xi\eta}\frac{t-s}{\tau},\tau\frac{t-s}{\tau})\nonumber \\
&\approx&\frac{1}{\eta^2}\int_0^t ds C(0,\tau\frac{t-s}{\tau})\nonumber \\
&=&\frac{k_B T}{\eta},
\end{eqnarray}
where we employed the same approximation, i.e., the expansion around the deterministic trajectory $y(x)$, and used the relation $\frac{\tau}{\xi}=\xi\cdot\sqrt{\pi}\eta\kappa/2k_B T$ in
 the Markovian limit.
The integral gives the diffusion coefficient.

In the above-mentioned short correlation regime, the Fokker-Planck equation becomes
\begin{equation}
\frac{\partial}{\partial t}P(x,t)=\frac{1}{\eta}\frac{\partial}{\partial x}\left(U'(x)(1-\kappa)+k_B T\frac{\partial}{\partial x}\right)P(x,t).\label{FP}
\end{equation}
It is one of our discoveries that the spatial correlation amounts to the systematic reduction of the drift velocity. The usual Langevin dynamics driven by the white noise is recovered in the case of
long-enough spatial correlation length which is expressed as $\kappa=0$.
Mathematically, this condition is satisfied when the ratio $\tau/\xi^2\rightarrow 0$, which only requires that
$\tau$ should vanish faster than $\xi^2$.
Intuitively, however, the spatial correlation length can be seen as infinite when it exceeds the typical length of the particles' displacement within the relaxation time $\tau$.
The steady state distribution is now given as
\begin{equation}
P^{{\rm st}}(x)={\it N} e^{-U(x)(1-\kappa)/k_B T},\label{steady}
\end{equation}
which has the form of canonical distribution with the renormalized potential $U(x)(1-\kappa)$ (Fig.3).
Here, the natural boundary condition is assumed, and ${\it N}$ is the normalization constant.

Let us now proceed to the escape rate problem characterized by the potential in Fig.1.
For the diffusion process governed by the Fokker-Planck equation (\ref{FP}), the mean first passage time is now
straightforwardly calculated, and its inverse gives the escape rate.
We assume the initial probability distribution $P(x,t=0)=\delta(x-x_0)$ confined in a metastable region bordered by
the free energy barrier at $x=x_{max}$ and the infinite wall at $x=-\infty$.
The mean first passage time $\tau(x_0)$ from the initial position $x_0$ is given by the standard procedure\cite{Risken} as a quadratic integral
\begin{equation}
\tau(x_0)=\frac{\eta}{k_B T}\int_{x_0}^{x_{max}}dy e^{U(y)(1-\kappa)/k_B T}\int_{-\infty}^y dz e^{-U(z)(1-\kappa)/k_B T},
\end{equation}
where the dependence on the initial point may not be important at low temperature.
For a high-enough barrier, the escape rate
 is evaluated by the steepest descent approximation at the maximum $x_{max}$ and the minimum $x_{min}$:
\begin{equation}
\Gamma\equiv\tau(x_0)^{-1}=\frac{1}{2\pi\eta}\sqrt{U''(x_{min})|U''(x_{max})|}(1-\kappa)e^{-\frac{\Delta U}{k_B T}(1-\kappa)}.
\label{escaperate}
\end{equation}

To confirm the validity of Eq. (\ref{escaperate}), we numerically solved the non-Markovian Langevin equation (\ref{basiceq}) in the special case of Landau potential $U=x^4/4-x^2/2$, under sufficiently short correlations of noise.
The spatio-temporal correlated noise can be given in a product form $f(x,t)=g_1(x,\xi)g_2(t,\tau)$ where the noises
$g_{i=1,2}(s,a)$ have the Gaussian correlation $\langle g_i(s,a)g_j(s',a)\rangle \propto \delta_{ij}e^{-\frac{(s-s')^2}{a^2}}$ with the correlation $a$.
In order to construct the colored noise $g_i(s,a)$, we considered an assembly of harmonic oscillators which mimics
the Langevin force
\begin{equation}
g_i(s,a)=\sum_k (x_k \cos\omega_k s+p_k\frac{\sin\omega_k s}{\omega_k}),
\end{equation}
where the coefficients $x_k$ and $p_k$ are the Gaussian stochastic variables with zero means and
the variances
\begin{eqnarray}
&&\langle x_k^2\rangle =\theta_i/m\omega_k^2\nonumber \\
&&\langle p_k^2\rangle =\theta_i/m,
\end{eqnarray}
with the mass $m=1$ and the noise strength $\theta$.
The density of the states $G_i(\omega)$ is chosen as
\begin{equation}
G_i(\omega)\propto\omega^2 e^{-\frac{\omega^2 a^2}{4}}.
\end{equation}
Then the desired Gaussian correlation with correlation length $a$ is achieved.
Note that the stochastic process $f(x(t),t)$ is not uniquely determined from the variance, and the product form is
one of the possible choices which guarantee the escape rate formula (\ref{escaperate}).
We fixed the parameters $\tau=0.01$, $\eta=1$, $k_B T=0.1$, and changed the correlation length $\xi$.
The correlation time $\tau$ is much shorter than the quantity $\eta\Delta U L_0^2/(k_B T)^2$ so that the correlation between external noise $f$ and
implicit noise contained in population $p$ is negligible. Here $L_0$ is the typical length scale accompanied by the probability distribution.
For each parameters, more than $200$ trajectories are simulated so that the mean-first-passage time well-converges.
The time step of the stochastic simulations is around $0.09$ which is longer than $\tau$, but sufficiently short so that the discretization of the equation makes sense.
The $\kappa$ dependence of the numerical escape rate $\Gamma$ is compared with both the theoretical prediction (\ref{escaperate}) and the traditional Kramers formula in Fig.2. We find a very nice agreement between the theoretical and numerical results.
\begin{figure}
\center{
\includegraphics[scale=0.8]{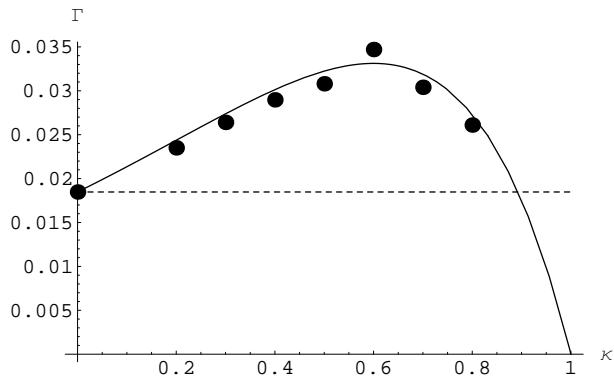}
}
\caption{The $\kappa$ dependence of the escape rate $\Gamma$ (\ref{escaperate})(the solid-line).
The broken-line shows the ordinary Kramers escape rate($\kappa=0$). 
The escape rate is maximized at a finite $\kappa$.
The Landau potential $x^4/4-x^2/2$ and parameters  $\tau=0.01$, $\eta=1$, and $k_B T=0.1$ are assumed.
The correlation length $\xi$ is varied which amounts to the change of $\kappa$.}
\label{activation}
\end{figure}
In Fig.3, we also examined the steady state distribution for parameters near the maximum of the escape rate of Fig.2 ($\tau=0.01$, $\xi=0.04750$, $\eta=1$, and $k_B T=0.1$).
\begin{figure}
\center{
\includegraphics[scale=0.8]{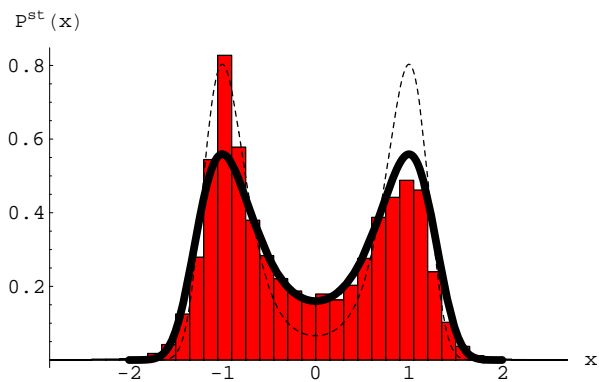}
}
\caption{The steady state distribution $P^{st}(x)$ for the Landau potential $x^4/4-x^2/2$ and parameters are $\tau=0.01$, $\xi=0.04750$, $\eta=1$, and $k_B T=0.1$.
The histogram is the result of the stochastic simulation which is composed of 30 samples of 2000 steps(transient duration is omitted).
The solid-line indicates the Eq.(\ref{steady}), and the broken-line is the canonical distribution
without potential renormalization.}
\label{activation2}
\end{figure}
In this way, the quasi-stable state becomes unstable due to the spatial randomness and leads to the thermal renormalisation  of both  the barrier-height and collisional prefactor in the Kramers escape rate.
Noting that the temperature is low enough, the normalization factor $(1-\kappa)$ is positive
and the escape rate shows a maximum at low but finite temperature.

In summary,
 stimulated by the work on the generalized Ornstein Uhlenbeck process in momentum space\cite{Arvedson},
we investigated the diffusion process with the spatially and temporally correlated noise for strong-damping regime.
A Fokker-Planck equation is derived in a kind of Markovian limit that is characterized by a dimensionless parameter $\kappa=2\tau k_B T/\sqrt{\pi}\xi^{2}\eta$.
The drift term is renormalized by the spatial randomness.
Intuitive understanding of this phenomenon should be as follows:
the spatial randomness yields domains of the spatial coherence of the random noise with the typical size $\xi$.
The systematic force pushes the particle out of the domain of coherence which amounts to an extra relaxation of the random noise.
Thus the systematic force $-U'(x)$ accompanied by the spatial relaxation
is equivalently replaced by the weaker force but without the spatial relaxation of the noise.
We think that the basic scenario is ubiquitous at least qualitatively.
The role of the potential renormalization is most clearly seen in the escape rate formula.
The consequent new escape rate
has a maximum at an optimal $\kappa$, due to a
competition between the reduction of the prefactor and that of the activation energy.

T.M. owes much to JSPS and is grateful to helpful discussions with Professor S.Tasaki and Professor P.Gaspard. A.S. and K. N. acknowledge partial support from JSPS.

\end{document}